\documentstyle[prd,aps,epsf]{revtex}

\def\be{\begin{eqnarray}}
\def\en{\end{eqnarray}}
\def\bea{\begin{eqnarray}}
\def\ena{\end{eqnarray}}

\begin{document}

\title{UPDATE ON GRAVITATIONAL-WAVE RESEARCH \footnote{To be published 
in the first volume of the ``Astrophysics Update", ed. J.W.Mason 
(Springer-Praxis,~2004) pp. 281-310}}
\author{L. P. Grishchuk}
\address{Department of Physics and Astronomy, Cardiff University, 
Cardiff CF24 3YB, United Kingdom \\
and  Sternberg Astronomical Institute, Moscow University, Moscow
119899, Russia\\ {\rm E-mail: grishchuk@astro.cf.ac.uk}}
\maketitle

\begin{abstract}
	The recently assembled laser-beam detectors of gravitational waves
are approaching the planned level of sensitivity. In the coming 1 - 2 years, 
we may be observing the rare but powerful events of inspiral and
merger of binary stellar-mass black holes. More likely, we will have to wait
for a few years longer, until the advanced detectors become operational.
Their sensitivity will be sufficient to meet the most cautious evaluations of 
the strength and event rates of astrophysical sources of gravitational waves.
The experimental and theoretical work related to the space-based laser-beam 
detectors is also actively pursued. The current gravitational wave research 
is broad and interesting. Experimental innovations, source modelling, methods 
of data analysis, theoretical issues of principle are being studied and 
developed at the same time. The race for direct detection of relatively 
high-frequency waves is accompanied by vigorous efforts to discover 
the very low-frequency relic gravitational waves through the measurements 
of the cosmic microwave background radiation. In this update, we will 
touch upon each of these directions of research, including the recent data 
from the Wilkinson Microwave Anisotropy Probe (WMAP). 
\end{abstract}

\section{Introduction}

	The concept of gravitational radiation has been with us for quite a 
long time. Thinking of the relativistic gravitational field in parallel with 
the familiar case of the electromagnetic field, it was natural to expect 
that there should exist waves of the gravitational field similar 
to the waves of the electromagnetic field. As Einstein \cite{ein} 
put it in 1913: ``The conviction had to come that Newton's law of 
gravitation is as incapable of desribing all gravitational phenomena 
as Coulomb's laws of electrostatics and magnetostatics are of 
electromagnetic phenomena". The decades of hard work have followed.      
In the beginning, the research was purely theoretical. Through doubts and 
controversies, the conceptual and mathematical issues of gravitational 
radiation have been clarified. Then, in the 60-s, the experimental work 
has started with the pioneering effort of J. Weber \cite{weber}. In the 90-s, 
gravitational waves have been observed indirectly via the measurement
of secular changes in the orbital parameters of the binary system of
neutron stars that includes the pulsar PSR 1913+16 \cite{taylor}. 
These days, gravitational waves are routinely taken into account in the
theoretical and observational studies ranging from orbital evolution of 
close pairs of compact stars to the early Universe cosmology. The
experimental progress has also been very impressive. 

We are now at a special and decisive p[oint. 
First, the relevance and importance of gravitational-wave research is fully
recognised by the communities of physicists and astronomers. The 
worldwide network of scientific collaborations has been 
established, reflecting the necessity of coincident observations at various
instruments and the need for joint analysis of the data \cite{igec},
\cite{lsc}. Second, the scientific runs have begun at the recently 
assembled sensitive 
laser-interferometric observatories - American LIGO \cite{ligo} 
and British-German GEO600 \cite{geo}. The French-Italian VIRGO \cite{virgo}
will be operational soon. Meanwhile, the Japanese TAMA300 \cite{tama}, along 
with the international network of bar detectors \cite{bars}, continue to 
collect data at their level of sensitivity. The design sensitivity 
of three LIGO interferometers plus VIRGO and GEO600 meets the realistic 
astrophysical predictions, so the direct detection of powerful (even if rare) 
sources, such as coalescing stellar-mass black holes, becomes 
likely. Third, the European Space Agency (ESA) and 
American NASA have agreed to share the costs of the space-based Laser 
Interferometer Space Antena (LISA) \cite{lisa}. LISA is planned to be launched 
around the year 2011, preceeded by a technology demonstration mission.
The plans for the advanced ground-based detectors, such as LIGO II, the
Japanese cryogenic laser interferometer LCGT, and, possibly, the European
EURO, are also maturing very quickly. These instruments of the next generation 
will detect a host of well anticipated sources, such as compact binary
stars, but really fundamental discoveries are also expected. Fourth, there
exists a strong competition on the side of purely astronomical 
means of indirect detection of gravitational waves. In fact, the anisotropy 
and polarisation measurements of the cosmic microwave background 
radiation (CMB) are likely to bring us decisive information on the 
fundamentally important relic gravitational waves much earlier than it 
will be done by direct methods. By all counts, gravitational-wave research 
is now one of the most exciting and promising areas of physical science.  

The weakness of gravity as a physical interaction is the strength of
gravitational waves as a tool of scientific research. It is difficult
to detect gravitational waves because they carry their energy practically
without scattering and absorption. But this is exactly because of this
difficulty that we have any chance to learn about what was happening at 
the beginning of the Universe, or in the depth of an exploding supernova,
or in the vicinity of, what we think are, merging black holes. It is also 
likely that the differing properties of gravitational waves will be a 
discriminating signature of different fundamental physical theories and 
modified gravities. 
 
In this review, we will start, in Sec. 2, with an elementary theory of 
gravitational waves. Then, in Sec. 3, we will discuss the current
status of gravitational-wave experiments. In Sec. 4 we will focus on
astrophysical sources of gravitational waves and the new physics that
will be learned from their observation. Sec. 5 is devoted to gravitational
waves and cosmology. In particular, some recent results from the WMAP 
will be discussed in the context of
gravitational-wave research, including the ``implications for inflation". 
Finally, we briefly summarise this update in Sec. 6. The background material
on the gravitational-wave research can be found in \cite{ll}, \cite{mtw}, 
\cite{wein}, and some of the previous reviews in \cite{thorne}, \cite{gr77}, 
\cite{schutz}, \cite{glpps}, \cite{cutlthorne}.

\section{Elementary theory of gravitational waves}

There are some similarities in the mathematical description of the 
gravitational field and the electromagnetic field. The electromagnetic field 
can be described by the components of the 4-vector potential 
$A^{\mu}$. The quantities $A^{\mu}$ are functions of time and spatial 
coordinates $x^{\alpha}$, $x^{\alpha} = (ct, x, y, z)$, and they obey 
the wave-like dynamical 
equations - the Maxwell equations. Other quantities, such as electric and 
magnetic components ${\bf E}$, ${\bf H}$, and the energy-momentum
tensor of the electromagnetic field, are calculable from $A^{\mu}$. The 
quantities $A^{\mu}$ allow the gauge freedom. This means that some 
seemingly different solutions of the field equations are in fact 
equivalent solutions in the sense of their physical manifestations.
Analogously, the gravitational field can be described by 10 components of 
the $4 \times 4$ symmetric tensor $h^{\mu \nu}$ as functions of $x^{\alpha}$. 
The quantities $h^{\mu \nu}$ obey the nonlinear wave-like dynamical equations -
the Einstein equations. The energy-momentum tensor of the gravitational
field $t^{\mu \nu}$ is calculable from the tensor $h^{\mu \nu}$. The quantities 
$h^{\mu \nu}$ allow the gauge freedom, so that some seemingly 
different solutions of the gravitational field equations describe in fact 
the physically equivalent configurations.

In contrast to the electromagnetic field, the gravitational field 
$h^{\mu \nu}$ is a nonlinear field. This means that a sum of two solutions 
of the field equations is not a new solution, and the gravitational
field, along with matter fields, is a source for itself. 
However, in a number of situations one can neglect the 
nonlinearity of the gravitational field. In this approximation we come
to the notion of weak gravitational fields and linearised gravitational 
waves.   

Many properties of linearised gravitational waves resemble those of 
electromagnetic waves. Gravitational waves propagate with the 
velocity of light $c$ and have two independent transvers polarization 
states. In its action on free masses, a gravitational wave (g.w.) 
exhibits some analogs of the electric and magnetic contributions of an 
electromagnetic wave (em.w.) acting on free electric charges. Gravitational 
wave field is dimensionless and its strength can be characterized by a 
dimensionless amplitude $h$. The amplitude $h$ decreases in the course of 
propagation from a localized source in inverse proportion to the traveled 
distance: $h \propto 1/r$. Gravitational waves carry away from a radiating 
system its energy, angular momentum and linear momentum. 

The linearised g.w. satisfy the wave equation
\begin{equation} 
\label{Lgw}
{h^{\mu\nu ,\alpha}}_{,\alpha} + 
\eta^{\mu\nu}{h^{\alpha\beta}}_{,\alpha ,\beta} - 
{h^{\nu\alpha ,\mu}}_{,\alpha}- {h^{\mu\alpha ,\nu}}_{,\alpha} = 0,
\end{equation} 
where the ordinary derivative is denoted by a comma and 
$\eta^{\mu \nu}$ is the metric tensor of the Minkowski space-time:
\begin{equation}
\label{Mi}
{\rm d} \sigma^2 = \eta_{\mu \nu} {\rm d}x^{\mu} {\rm d} x^{\nu} =
c^2{\rm d}t^2 - {\rm d}x^2 - {\rm d}y^2 - {\rm d}z^2.   
\end{equation} 
The first term in Eq. (\ref{Lgw}) is the familiar d'Alembert (wave) operator.
A plane-wave solution to Eq.~(\ref{Lgw}) is given by 
\begin{equation} 
\label{plw}
h^{\mu\nu}= a^{\mu\nu} e^{i  k_{\alpha} x^{\alpha}}, 
\end{equation} 
where $k_{\alpha}k^{\alpha} = 0$, reflecting the fact that a g.w. 
propagates with the velocity of light. Because of this condition, the field
equations (\ref{Lgw}) require the 10 components of the constant matrix 
$a^{\mu\nu}$ to satisfy 4 constraints: $a^{\mu\nu}k_{\nu} = 0$.
Then, the quantities $a^{00}$ and $a^{0i}$ can be expressed in terms of 6 
components of the matrix $a^{ij}$. This matrix itself includes the part
$\tilde{a}^{ij}$ satisfying the further 4 constraints:  
$\tilde{a}^{ij}k_j = 0, ~ \tilde{a}^{ij}\eta_{ij} = 0$. These
remaining 2 degrees of freedom (sometimes called the TT-components) 
fully determine the observational manifestations of the plane 
wave and its energy-momentum characteristics. Indeed, it is easy to show
that the gravitational energy-momentum tensor 
\begin{equation} 
\label{gwtmn}
t_{\mu\nu} =\frac{c^4}{32 \pi G}\left[{h^{\alpha\beta}}_{, \mu}
h_{\alpha\beta , \nu} - \frac{1}{2} h_{, \mu}h_{, \nu} \right]
\end{equation}  
depends only on the TT-components of the field:
\begin{equation} 
\label{gwenm}
t_{\mu\nu}= \frac{c^4}{32 \pi G} k_\mu k_\nu \left[2 \tilde{a}^{ij} 
\tilde{a}^{*}_{ij}\right],
\end{equation} 
where we have dropped (as we normally do in the case of electromagnetic 
waves) the purely oscillatory terms.

How does a gravitational wave affect the free (i. e. not subject to any 
other forces or constraints) masses ? In the case of an em.w.,
we could have noticed the displacement of a charged particle with respect to a 
neutral particle placed initially in the same point. In the case of a
gravitational wave, there is no particles neutral to the gravitational
interaction, so we need to study the relative displacement of particles 
separated initially. This brings us to the analysis of tidal effects of 
a gravitational wave, similar to the tidal effects of a Newtonian 
gravitational field. Let one of free masses define the origin of our
coordinate system, $x_{(1)}^i =0$, whereas the second mass is placed initially 
at $x_{(2)}^i = l^i$ and its (small) displacement caused by the gravitational
wave is denoted by $\xi^i$. Then the equations of motion of the second mass
are: 
\begin{equation} 
\frac{d^2\xi^j}{dt^2}=\frac{1}{2}\omega^2 \tilde{a}^j_k l^k e^{i\omega t}.
\label{mrelmotws}
\end{equation} 
These equations depend only on the TT-components of the wave. 

For a wave propagating, say, in $z$-direction it is convenient to write
the TT-components explicitely:
\begin{equation} 
\label{zwave}
h_{xx}= -h_{yy} = h_{+} \sin (\omega t -kz + \psi), ~~~h_{xy} =
h_{\times} \cos (\omega t - kz + \psi).
\end{equation} 
Then, the relevant solution to Eq. (\ref{mrelmotws}) reads
\begin{equation}  
\label{sol}
x = l_1 - \frac{1}{2}l_1h_{+}\sin(\omega t +\psi) -\frac{1}{2}l_2h_{\times}
\cos(\omega t + \psi),~~ 
y =l_2 - \frac{1}{2}l_1h_{\times}\cos(\omega t +\psi) +
\frac{1}{2}l_2h_{+} \sin(\omega t + \psi),~~ z= l_3.
\end{equation}      

\begin{figure}[tbh]
\centerline{\epsfxsize=8cm \epsfbox{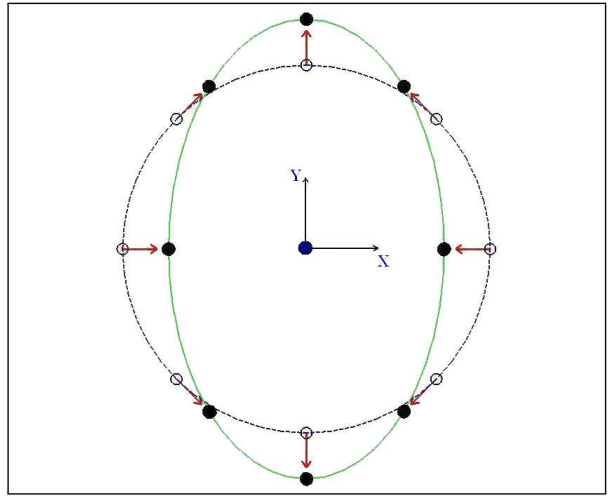}}
\caption{Motion of free particles in the field of a linearly-polarised
gravitational wave}
\label{Fig.1}
\end{figure}

The masses that lie (on average) on the ring 
$(l_1)^2 + (l_2)^2 =l^2, l_3=0$, oscillate 
around their ring positions. In Fig. 1 we show a quarter of the cycle
caused by a wave with the $+$ polarization state (i.e. when 
$h_{+} \neq 0, ~ h_{\times} =0$). The pattern of oscillations enforced by 
the $\times$ polarization state (i.e. when $h_{\times} \neq 0, ~ h_{+} = 0$) 
can be obtained from Fig. 1 by its rotation by the 
angle $45^o$ in the $(x, y)$ plane. The general motion of a particle 
is a linear superposition of these oscillations, as seen in Eq. (\ref{sol}). 
The oscillatory deformation
of a sphere of masses, surrounding the central mass at the origin, is 
desribed by a liner combination of spherical harmonics 
$Y_{lm}(\theta, \phi)$ with $l=2$ and $m= \pm 2$, and where the polar 
axis is taken along the $z$ direction.

Equations (\ref{sol}) suggest that the motion of free particles is 
confined strictly to the planes $z =const.$ This conclusion comes about only 
because we have so far neglected the smaller terms, containing the
products of $h_{+},~h_{\times}$ with the extra small factor $l/\lambda$, 
where $l$ is the separation between masses and 
$\lambda =2 \pi c/\omega$ is the gravitational wavelength. When these
terms are taken into account, the perturbed positions $x, y$ and, most
importantly, $z$ receive further contributions. The gravitational wave 
drives the masses not only in the plane of the wave-front, but also, to a 
smaller extent, back and forth in the propagation direction \cite{gr77}. This 
extra component of motion is similar to the one caused by the magnetic field 
and the Lorentz force of the em.w. acting on a charged
particle. In Fig. 2 we show typical displacements of the masses 
forming (on average) the plane $z=0$, when the ``magnetic" 
contribution to their motion is also taken into account. It is seen from
this figure that the entire plane of particles is being bent in an oscillatory 
fashion. This ``magnetic" component of motion will play a certain
role in the analysis of observations with laser interferometers \cite{bg}.

\begin{figure}[tbh]
\centerline{\epsfxsize=8cm \epsfbox{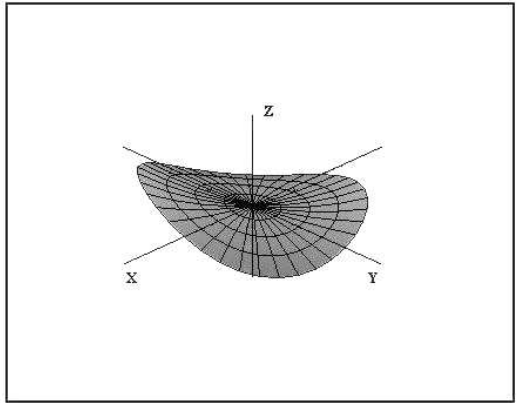}}
\caption{Motion of free particles with the ``magnetic" component taken into 
account}
\label{Fig.2}
\end{figure}
 
The amplitudes $h_{+}, h_{\times}$ are determined by the source of 
the gravitational waves. To find the amplitudes, we should 
replace the zero in the right hand side of Eq. (\ref{Lgw})
by the source term $(16 \pi G/ c^4) T^{\mu\nu}$, where $T^{\mu\nu}$ is the
energy-momentum tensor of the source, and seek the retarded solutions
to the wave equation. Assuming that the distance $R_0$ to the source is 
much larger than the wavelength, one can write
\be
\label{atild2}
h_{+}(k_{\alpha}) = \frac{1}{R_0} \frac{2 G}{c^4} [\hat{T}^{11}(k_{\alpha})-
\hat{T}^{22}(k_{\alpha})],~~~ h_{\times}(k_{\alpha}) = 
\frac{1}{R_0} \frac{4 G}{c^4} \hat{T}^{12}(k_{\alpha}), 
\en
where the Fourier components $\hat{T}^{\mu\nu}(k_{\alpha})$ 
are defined by
\be
\hat{T}^{\mu\nu}(k_{\alpha})=\int\left( \int T^{\mu\nu}(t_r,r_0) d^3 r_0\right)
e^{-ik_{\alpha}x^{\alpha}} d^4x  
\nonumber 
\en
and $t_r$ is retarded time. These formulas allow one to 
evaluate the typical amplitude $h$ from a given source:
\be
\label{ampl}
h \sim \frac{1}{R_0} \frac{G M}{c^2} \left(\frac{v}{c}\right)^2,
\en
where $M$ is the total mass of the source and $v$ is the characteristic 
non-spherical velocity of the matter bulk motion. For a bound system
like a binary star, $(v/c)^2 \sim GM/c^2 a$, where $a$ is the size of
the system. An accurate calculation for a binary in a circular orbit, 
consisting of 
masses $M_1$, $M_2$ separated by the distance $a$, and after averaging
over the orbital period and orientation of the orbital plane, gives
\be
\label{amplb}
h = \left( \langle h_{+}^2 \rangle + \langle h_{\times}^2\rangle \right)^{1/2}=
\left( \frac{32}{5}\right)^{1/2} \frac{1}{R_0} \frac{G^{5/3}}{c^4}
\frac{M_1 M_2}{(M_1 +M_2)^{1/3}} (\pi f)^{2/3},
\en
where the emitted g.w. frequency $f$ (in $Hz$) is 
\be
f = \frac{1}{\pi} \left[ \frac{G(M_1 + M_2)}{a^3}\right]^{1/2}.
\en
Note that the amplitude $h$ depends on a particular combination of masses: 
\be
\label{chirp}
\frac{M_1 M_2}{(M_1 +M_2)^{1/3}} = {\cal M}^{5/3}, ~~~{\rm where}~~~
{\cal M}= \frac{(M_1M_2)^{3/5}}{(M_1+M_2)^{1/5}},
\en 
so that $h \propto (1/R_0) {\cal M}^{5/3} f^{2/3}$ and ${\cal M}$ is 
sometimes called a chirp mass. 
 
\section{Current status of gravitational wave detectors} 

The possible methods of detecting gravitational waves, as well as the 
associated difficulties, can be seen from the discussion above. Fig.1 
is helpful for understanding the principles of mechanically coupled 
detectors (bar detectors) and electromagnetically 
coupled detectors (laser interferometers). A bar 
detector is essentially a mechanical oscillator consisting of two masses 
connected by a spring. As an illustration, one can think of two elastically 
connected masses lying, say, at the opposite ends of the $x$ axis in Fig. 1. 
The size of bar detectors is normally small, a couple of meters or so. 
A bar detector is a relatively narrow-band instrument. It is mostly 
sensitive to g.w. frequencies in the vicinity of the main eigen-frequency 
of the bar. In the presently operating instruments the resonant frequency
is around $\sim 1 kHz$. There are some advantages in using spherically shaped 
mechanical detectors. The construction of such detectors is currently 
taking place (see, for example, \cite{spheres}). While the operating 
bar detectors continue to collect useful information \cite{astone}, 
\cite{finn}, \cite{astone2}, we will concentrate on laser interferometers.    

The laser interferometer technique is based on free masses-mirrors whose 
relative distances are monitored by bouncing light. A ground-based 
interferometer is normally an $L$-shaped configuration. The light 
beamsplitter is in the corner of $L$, while the reflecting mirrors are at 
the ends of the configuration and near the beam splitter. The central mass 
in Fig.1 is an illustration of the beamsplitter and corner mirrors,
whereas the end mirrors are located, say, in the positive directions of 
$x$ and $y$. Obviously, mirrors in real interferometers are not free
masses, they are suspended like pendulums. But they behave essentially as free
masses in their motion along the corresponding arm. The 
length of the arms of an interferometer can be large. For instance, it 
ranges from 600 meters in GEO600 to $4 km$ in largest of LIGO 
interferometers. In contrast to bar detectors, laser interferometers are 
relatively broad-band instruments. They are sensitive to frequencies
in the interval $\approx (30 - 10^4) Hz$. Fig.1 shows the simplest case of 
a single monochromatic wave with one polarisation state arriving from
the orthogonal direction to the detector's plane, but the 
response of the detector to the general case of the incoming wave is 
also calculable. 

It is clear from Eq. (\ref{sol}) that the relative displacement of
free masses is proportional to the incoming wave amplitude $h$:
$\delta l/l \approx h$. Which numerical values of $h$ can be expected from 
astrophysical sources in the surrounding Universe~? Consider one of the 
most powerful and efficient emmitters - 
a pair of compact stars orbiting each other at a tight orbit. Specifically, 
consider two neutron stars with masses $M_1 = M_2 = 1.5 M_\odot$ at the 
late stage of their inspiral. Orbiting each other at separation $a = 100 km$,
they emit gravitational waves at frequency $f = 200 Hz.$ 
The emitted intensity of radiation is very high by astronomical standards:
$3 \times 10^{52} erg/sec$. Let the distance to the source be $R_0=100 Mpc$.
We cannot take this distance much shorter than $100 Mpc$, because the expected 
number of such events per year in a smaller volume of the surrounding 
Universe would be 
less than 1. Then, formula (\ref{amplb}) says that the amplitude at Earth is 
$h \approx 10^{-22}$. This is an increadibly small number. It enters any
conceivable method of detection of gravitational waves and explains why it 
is so difficult to observe them. In a $4km$ long interferometer, we need 
to beat all the noises and measure the mirror's displacements at the level  
of $4 \times 10 ^{-17} cm$.       

The interferometer monitors the time-dependent difference of the relative 
distance variations in the two arms. Using the terminology of solid state 
physics, this quantity is called the (dimensionless) strain. Regardless 
of the presence or absense of the useful g.w. signal, there will always 
be some strain noise $n(t)$ in the output of the detector. The mean-square 
value of the noise can be expressed as an integral over the noise 
power spectral density $S_n(f)$:
\be
\label{powsd}
\overline{n^2(t)} = 2 \int_0^{\infty} S_n(f) df. 
\en
The square root of $S_n(f)$ is the noise amplitude $\sqrt{S_n(f)}$.
This quantity has the dimensionality of ${Hz}^{-1/2}$, and it is 
this quantity that is usually plotted on the sensitivity graphs. 
In Fig. 3 (taken from \cite{ligo}) we show the recent status of the noise 
amplitudes in the $4km$ interferometer of the LIGO Livingston cite. 
One can see the great progress that has been made on the way of 
reaching the design goal of sensitivity, shown by the solid line. 
The lower frequency part of the sensitivity 
curve is dominated by seismic perturbations, the central part by thermal 
noise in mirror's suspensions, and the higher frequency part by the shot 
noise of the laser light. The advanced LIGO sensitivity curve will be lower
than the solid line In Fig.3 by a factor of 10 across all the frequencies.
This upgrade of LIGO is planned to take place in about the year 2007. The 
advanced instruments will also allow optical configurations
in which the sensitivity can be increased in a certain narrow frequency
band at the expense of lowering the sensitivity outside the chosen band. 

To compare qualitatively the astrophysical signal $S$, represented
by the dimensionless amplitude $h$, with the detector's r.m.s. noise $N$, 
we should calculate, as formula (\ref{powsd}) suggests, the product 
$\sqrt{S_n(f)} \sqrt{\Delta f}$, where $\Delta f$ is the appropriate
bandwidth in the noise spectrum. Taking the initial LIGO design sensitivity 
$3 \times 10^{-23} Hz^{-1/2}$ at $f= 200 Hz$ and $\Delta f = f$, we would 
get $\sqrt{S_n(f)} \sqrt{\Delta f} = N = 5 \times 10^{-22}$.
This $N$ is a factor of 5 higher than the signal $S= 10 ^{-22}$ from
the coalescing neutron stars considered above. The ratio 
$S/N = 1/5$ would suggest that, by a significant margin, the signal is 
not measurable. The reality, however, is somewhat better than this estimate. 
One should take into account  the apriori knowledge of the expected
waveform. If the waveform is known, one can use the well developed technique 
of matched filtering. This method recovers the signal by, 
effectively, reducing the appropriate $\Delta f$ and the relevant 
amount of detector's noise. For a quasi-periodic signal, this is 
achieved by using a long observation time T: $1/ T \sim \Delta f \ll f$.
In the case of a coalescing binary, the dominant g.w. frequency is 
increasing with time, but the binary still executes almost
$n=200$ cycles before changing its frequency by a factor of 2. This 
points out to a possible increase of the estimated $S/N$ by a factor 
$\sqrt{n}$. The importance of accurate modelling of the expected 
waveforms is well appreciated. Presently, this is an important challenge 
for theorists (see, for example, \cite{blanch}).  

\begin{figure}[tbh]
\centerline{\epsfxsize=9cm \epsfbox{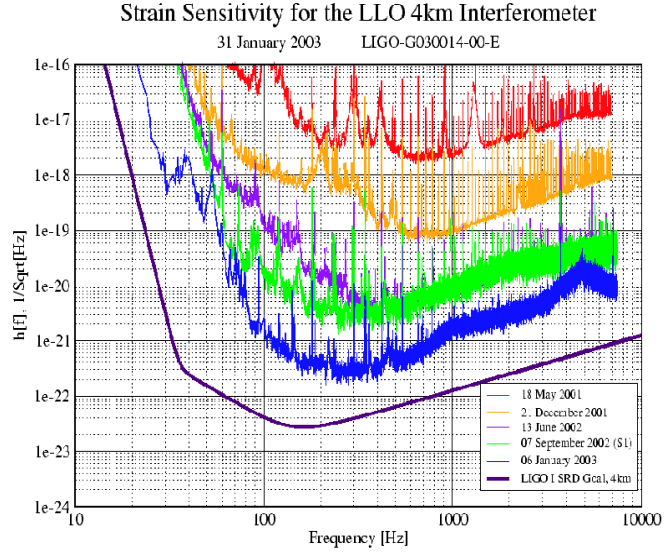}}
\caption{The status of sensitivity of one of the LIGO interferometers}  
\label{Fig.3}
\end{figure}

Fig. 4 (taken from \cite{glpps}) shows the accurately calculated  
$S/N$ ratios for the initial laser interferometers. As expected,
the $S/N$ grows with the total mass $M$ of the coalescing binary, because 
the signal gets larger. However, binaries with the total mass greater 
than about $80 M_\odot$ radiate at too low frequencies before merging. At
these low frequencies the detector noise is high, as seen in Fig. 3. 
Therefore, the $S/N$ decreases again. Certainly, the equal mass binaries
with the total mass greater than $M \sim 10 M_{\odot}$ can only be pairs
of black holes, according to our present understanding of stars and stellar
evolution.
 
The LIGO-GEO collaboration has reported \cite{barish} the first upper limits
on coalescing binary systems, as well as other possible g.w. sources. 
These limits were derived from observations at the currently achieved 
sensitivity. They are not yet significant from the astrophysical point of 
view, but, outside the frequency interval probed by the bar detectors, 
they are tighter than other upper limits experimentally established 
so far. The necessary steps for increasing the sensitivity in the initial 
and advanced ground-based interferometers are well recognised and are
being taken up. 

The space-based interferometer LISA will be sensitive to g.w. in the interval 
$10^{-4} - 10^{0} Hz$, that is, to lower g.w. frequencies as compared 
with ground-based instruments. Some new types of astronomical sources
will be accessible to this detector. LISA will consist of three 
spacecraft, forming an equilateral triangle of side 5 million km, in 
a heliocentric orbit, lagging behind the Earth by $20^o$. The phase shifts of
the laser light traveing along all three sides of the triangle will 
monitor the light-travel distance between the small drag-free test masses 
inside the spacecraft. The nominal lifetime of the mission is 5 years. 
Fig.5  shows the LISA design sensitivity curve and some of 
interesting sources of gravitational waves. This single figure attempts 
to show the noise curve of the entire assembly of spacecraft together 
with the strengths of g.w. sources of 
different nature - from quasi-periodic to stochastic - and therefore 
the figure should be treated with some care. The instrumental noise 
level is shown here in bins of $3 \times 10^{-8} Hz$, which are appropriate 
for a 1 year integration time. In other words, the r.m.s. instrumental
noise $\sqrt{S_n(f)}$ in units of $(Hz)^{-1/2}$ (see Eq.(\ref{powsd}))
is multiplied with $\sqrt{3 \times 10^{-8} Hz}$ at each frequency,
thus producing a dimensionless spectral quantity. (For the latest 
amendments of the LISA noise curve  see, for example, \cite{lisanoise}.) 
This is done mainly in order to emphasize that during this observational
time the g.w. signals from many of the galactic white-dwarf binaries can 
be resolved and removed from the data. This means that their g.w. noise will
not prevent us from seeing something much more interesting - a stochastic
background of relic gravitational waves. The sharp drop of signal from
the galactic binary white dwarfs, shown in Fig.5 at 
$f =2 \times 10^{-3} Hz$, illustrates this assumed operation with the data
\cite{glpps}, \cite{cornish}, \cite{tinto}, and not the total lack
of galactic binaries radiating at frequencies higher than 
$2 \times 10^{-3} Hz$. The continuation of this curve at 
$f > 2 \times 10^{-3} Hz$ shows the much smaller 
g.w. noise from unresolved extragalactic white dwarf binaries. In accord 
with this way of describing the dimensionless instrumental noise, the 
dimensionless g.w. signal amplitudes are also calculated in bins of 
$3 \times 10^{-8} Hz$ around any given frequency $f$. In the next
section, we will present more details on astrophysical sources for
ground-based and space-based instruments.

\begin{figure}[tbh]
\centerline{\epsfxsize=8cm \epsfbox{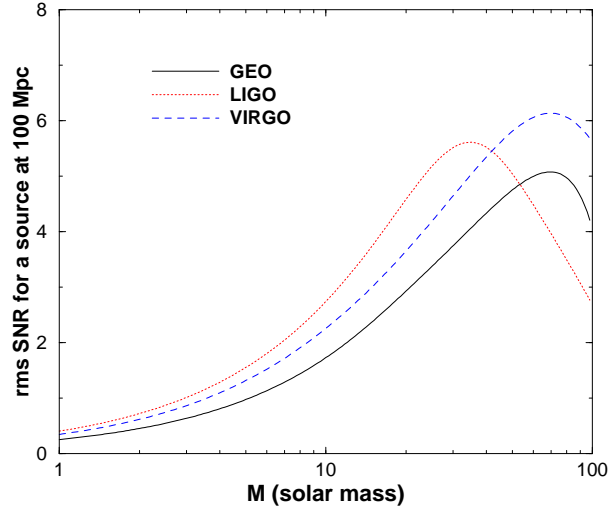}}
\caption{Signal to noise ratio in initial interferometers as a function
of total mass for inspiral signal from binaries of equal masses and averaged
over source inclination}
\label{Fig.4}
\end{figure}

\section{Gravitational waves and astrophysics}

It is common to divide the sources in groups, depending on whether
they are accessible to ground-based or space-based instruments. It is
also common to call them, respectively, the high-frequency and low-frequency
sources. 

\subsection{High-frequency sources}

We will focus on the sources that are likely to be the first sources 
detected by the ground-based instruments. It is clear from Eq. (\ref{ampl}) 
that a powerful source of gravitational waves should involve large masses and
relativistic velocities. A pair of neutron stars inspiraling toward each
other under the influence of the g.w. radiation reaction force is a
primary example. In the last few seconds of inspiral, before the neutron 
stars ``touch" each other and merge, the frequency of gravitational 
radiation increases from $200 Hz$ to about $1200 Hz$. 
The orbital velocity increases from $v/c \approx 0.2$ to $v/c \approx 0.4$.
In the last few seconds the binary emits $1.1 \times 10^{53} ergs$ of energy
in the form of gravitational waves. This is a 2$\%$ of its total rest-mass 
energy $Mc^2$. What can be more powerful and efficient a source of 
gravitational radiation than this one ? Only a pair of even more massive and 
even more compact stars. Astronomers call them stellar-mass black holes.  
 
If the coalescence of a binary neutron star (NS+NS) or a binary black
hole (BH+BH) happened in our Galaxy, it would be easily detectable 
by the already operating instruments. The problem is that these catastrophic 
events are expected to take place only once per a very long while. In 
order to have a reasonable chance of seeing, say, 3 events per year, we 
have to survey a large volume of space, which includes many galaxies. 
This means that the instrument's sensitivity should be so high, that the 
events could be seen from the edges of this large volume. This 
consideration explains the importance of theoretical evaluations of
the event rates in a typical galaxy. 

The event rate of coalescing NS+NS systems is partially constrained by 
pulsar observations in our Galaxy. Three NS+NS binaries are known, each 
involving a pulsar, whose coalescence time is less than the Hubble time and is,
on average, $3 \times 10^8$ years. Starting with these 3 binaries, one would
evaluate the NS+NS event rate as 1 per 100 million years. However, we observe 
only about 1$\%$ of the galactic volume, so the coalescence rate can easily
be raised to $10^{-6} yr^{-1}$ \cite{phinney}. Very likely, the event rate  
for NS+NS systems is significantly higher than this estimate, if only 
because of the fact that not all NS+NS systems include a currently 
observable pulsar. The observational situation with black holes is more 
uncertain. There are a dozen of BH candidates in X-ray binary systems, but 
they all are in pairs with non-degenerate companions. So far, there is no 
observational evidence for NS+BH or BH+BH binaries. Nevertheless, one can 
make some evaluations on the basis of the star formation rates. 

It is believed that the neutron star progenitors have masses 
greater than $10 M_\odot$, whereas the black hole progenitors have masses
greater than $80 M_\odot$. The Salpeter function for the star formation rate is
$$
\frac{dN}{dt d(M/M_\odot)} \simeq 
\left(\frac{M}{M_\odot}\right)^{-2.35} yr^{-1}.   
$$
Integrating this function over $M$ and using the lower limit of integration, 
one finds the ratio of the expected numbers of progenitors: 
$$
\frac{N(M> 80M_\odot)}{N(M> 10M_\odot)}=
\left(\frac{80M_\odot}{10 M_\odot}\right)^{-1.35} \simeq 0.06. 
$$    
It is reasonable to think that, despite all the complexities and
differences in binary evolution, the ratio of coalescence rates will
also be given by approximately the same quantity,
$$
\frac{{\cal R}_{BH}}{{\cal R}_{NS}} =
\left(\frac{80M_\odot}{10 M_\odot}\right)^{-1.35} \simeq 0.06. 
$$
This expectation turns out to be in rough agreement with the results
of detailed numerical population synthesis calculations. 

Numerical calculations take into account all the available observational 
information. Their advantage is in that one can follow
not only the channels leading to the NS+NS, NS+BH, BH+BH systems,  
most interesting for gravitational-wave astronomy, but also other
evolutionary outcomes, which allow comparison with observations in their
own right. The population synthesis results cannot be less reliable than 
purely ``observational" estimates, as they are controlled by the same
available observational data. Discrepancies in final results 
exist because of uncertainties in astrophysics, not because one 
of the methods is inherently less reliable than another. 
This is especially true with regard to the
NS+BH and BH+BH binaries, where the current purely ``observational"
evaluations would have to begin with zero. 

The results of conservative population synthesis calculations
\cite{lpp}, \cite{glpps} have been checked on their 
consistency with other evolutionary outcomes. 
The calculations show that the NS+NS rate is expected to be at the level 
${\cal R}_{NS} = 3 \times 10^{-5} yr^{-1}$, whereas the BH+BH rate is at 
least one order of magnitude lower. For further estimates we will 
take it at the level ${\cal R}_{BH} = 0.06~{\cal R}_{NS} = 
2 \times 10^{-6} yr^{-1}$.
These rates for a typical galaxy, ${\cal R}_G$, determine the rates for a
given cosmological volume, ${\cal R}_V$, which includes many galaxies.
When deriving the ${\cal R}_V$, it is convenient to use a conservative 
estimate for the baryon content of the Universe. This brings us to the
relationship   
\[
{\cal R}_V \approx 0.1 {\cal R}_{G} \left( \frac{r}{1 ~Mpc}\right)^3.
\]
Thus, it is expected that within the volume of radius $r=100~Mpc$ and
during 1 year, there will be 3 of NS+NS events and only 0.2 of
NS+BH or BH+BH events. The increase of the radius to $r= 200~Mpc$
increases the volume and the event rates by a factor 8.
It is interesting to note that, despite all the diversity of approaches
in the literature, there exists some tendency to convergence of the  
final results \cite{knst}, \cite{kkl}, \cite{sipsig}, \cite{rasio},
\cite{sig}.   

The derived event rates are the basis for the calculation of the expected
detection rates. The important fact is that the mass of a typical 
neutron star is $1.4~M_\odot$, whereas the mass of a typical black hole
is $(10 - 15)~M_\odot$. The averaged mass of the observed
black hole candidates in the X-ray binaries is $M_{BH} \simeq 8.5 M_\odot$.
A pair of black holes is a more powerful source of gravitational waves 
than a pair of neutron stars, and therefore black hole binaries
can be seen by a given instrument from much larger distances. 
When it comes to the detection rates, the lower event rate of BH+BH 
sources is more than compensated by their larger masses. Indeed, the 
optimal signal to noise ratio is \cite{thorne}, \cite{fh}, \cite{glpps}: 
$S/N \propto {\cal M}^{5/6}/r$. At a fixed $S/N$, the detection volume is
proportional to $r^3$ and therefore to ${\cal M}^{5/2}$. The detection rate
${\cal D}$ for binaries of a given class is the product of their
event rate ${\cal R}_V$ and the detector's registration volume 
$\propto {\cal M}^{5/2}$ for these binaries. Therefore, one obtains
\[
\frac{{\cal D}_{BH}}{{\cal D}_{NS}} =    
\frac{{\cal R}_{BH}}{{\cal R}_{NS}}\left(\frac{{\cal M}_{BH}}{{\cal M}_{NS}}
\right)^{5/2}= 0.06 \left(\frac{8.5 M_\odot}{1.4 M_\odot}\right)^{5/2}
\simeq 5.5.
\]
This remarkable result is consistent with the more accurate numerical 
calculation of $S/N$ displayed in Fig. 4. 

It is seen from Fig.4 that g.w. signals from NS+NS at $r = 100 Mpc$ cannot 
be regarded detectable by initial interferometers. The situation is much
better for havier pairs of NS+BH and BH+BH. If the total mass of a
BH+BH binary is near $(20 - 30) M_\odot$, then $S/N \approx 2$ even if
the binary is placed at $r = 200 Mpc$. It follows from the event
rate ${\cal R}_{BH}$ discussed above, that in this larger volume one expects 
a couple of BH+BH events per year. The simultaneous observations on two or
three instruments will significantly diminish the probability of false 
alarms to such events. This is why it is argued in Ref. \cite{glpps} that the
coalescing black holes will probably be the first sources detected by
the initial ground-based interferometers, when they reach their planned 
sensitivity. Of course, the discussed estimates are statistical 
by the very nature of things, and they have significant systematic 
unceratinties. It will not be very surprising if the reality is somewhat 
better or somewhat worse than what the mean values suggest. It is 
important, however, that even the most pessimistic evaluations of the 
NS+NS and BH+BH rates indicate that there should be many detections per 
year in the advanced interferometers.
 
If the coalescing black holes are detected first, it is likely that at 
the beginning we will only have a proof that the objects are black holes 
in astronomical sense - an inspiraling pair of heavy compact masses. 
The next step will be to try to understand their real physical nature. The
general-relativistic black holes possess event horizons, which are
supposed to merge into the resulting black hole, which will then emit a 
damped train of ringdown waves at specific frequencies, and
so on \cite{fh}, \cite{bd}. This fascinating physics will be testable when 
the good quality data are available. This analysis will also require 
the continuation of the intense effort 
on the side of analytical and numerical calculations (for a recent 
review of numerical relativity, see \cite{lehner}). 
 
Direct detection of first sources will be not the end, but only the 
beginning of the
observational g.w. astronomy. In the long run, the aim of g.w. science
is to explore a great variety of sources, many of which can 
hardly be seen in electromagnetic radiation. Needless to say that there is
also a great chance of discovering new and totally unexpected sources.
In addition to coalescing binary stars, many other sources will 
eventually be detected. For instance, the long-recognised importance of 
the core collapse of massive stars \cite{rrw}, \cite{erdl} as g.w. sources 
has been reinforced by the mounting evidence of asymmetries during the 
supernova explosions, by the likelihood of forming and quick 
collapse of very massive early stars, by the association of 
supernovae events with gamma-ray bursts, etc. (For a recent review and 
extensive list of references, see for example \cite{new}.) The 
tidal disruption of a neutron star by its companion, the various sorts 
of stellar instabilities \cite{and}, the slightly deformed spinning 
neutron stars, young pulsars, and low-mass X-ray binaries \cite{pss}, 
\cite{bccs} - are also the astrophysically important and interesting 
g.w. sources that will be studied, very likely, by advanced detectors.

\subsection{Low-frequency sources}  

It is common to call the low-frequency g.w. sources as ``sources for LISA".
Some of them are displayed in Fig. 5. As explained in the previous section,
the dashed line shows the g.w. confusion noise from the binary 
white dwarfs, WD+WD, mostly concentrated in the disk of our Galaxy. LISA 
will not only detect thousands of WD+WD systems radiating at 
$f > 2 \times 10^{-3}Hz$, but is capable of doing this so 
accurately that their contributions can be removed individualy 
from the data. Some known binaries consisting of
degenerate and normal stars will also be detectable. Their angular 
coordinates and distance can be measured with high precision \cite{cv}. 
In fact, the well identified galactic binaries serve as guaranteed 
sources for LISA, and they will help to test LISA's performance.   

\begin{figure}[tbh]
\centerline{\epsfxsize=9cm \epsfbox{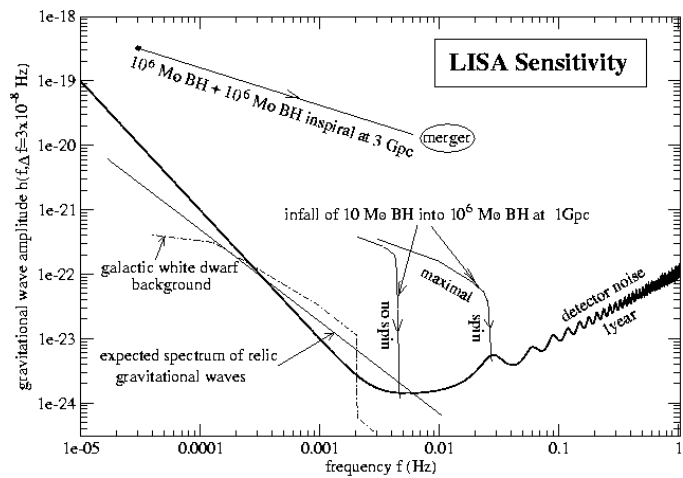}}
\caption{Some interesting sources in comparison with LISA sensitivity}
\label{Fig.5}
\end{figure}

There is growing evidence for the existence of binary supermassive black 
holes (SMBH, $M \ge 10^6 M_\odot$) in the centers of merging galaxies.  
(For a recent review of SMBH formation see \cite{shapiro}). 
Certainly, a coalescing pair of SMBH is an extremely bright g.w. emitter.
These sources are located at cosmological distances, so that the red shift
of the incoming gravitational waves becomes important. But if the total
mass of the pair is not significantly larger than $10^{7} M_\odot$, the
coalescing pair can still be visible by LISA at frequencies up to 
$f \approx 10^{-4} Hz$, even if the source is located at the red 
shift $z=3$. The upper line in Fig.5 shows the effective gravitational 
wave amplitude, relative to the plotted LISA instrumental noise, for a 
source consisting of two SMBH. The dot shows the signal when the 
inspiraling pair was only 1 year prior to merger. The presence of spins
of SMBH changes the LISA signal \cite{vecchio}. The trouble with these 
super-powerful 
sources is that their event rate is very uncertain. One of difficult
issues is whether a SMBH forms by direct collapse of gas in deep
galactic potential wells or by hierarchical build-up of pre-galactic 
structures. Nevertheless, it is estimated that LISA may see 0.1-1 events
per year, or maybe a factor of 10 more \cite{kh}.    

It seems certain that, from time to time, a SMBH will be closely 
approached by nearby compact stars - white dwarfs, neutron stars and 
stellar-mass black holes. A compact star orbiting a SMBH is a 
powerful g.w. source. However its orbit and the emitted waveform depend
strongly on the spin of SMBH. Fig.5 shows two curves (taken from
\cite{cutlthorne}) describing the performance of a $10 M_\odot$ BH 
completing its inspiral toward a $10^6 M_\odot$ SMBH. The left curve
corresponds to a non-spinning SMBH and a circular orbit of a 
stellar-mass BH. The right curve 
describes the signal from a prograde, circular, equatorial orbit 
around a nearly maximally spinning Kerr SMBH. Both curves begin at 
frequencies when the time to merger is 1 year. A serious theoretical 
problem is the construction of reliable templates for these 
sources. One should expect that a typical orbit will be eccentric, 
non-equatorial, and subject to the radiation-reaction force corrections. 
Correspondingly, the waveforms could be extremely complicated. Some 
progress in this area is reported in \cite{ss}. 

Finally, Fig.5 shows the expected level of relic gravitational waves.
This is a fundamentally important signal from the very early Universe.
Its explanation requires cosmological notions and some elements of quantum 
physics. Since the relic gravitational waves and primordial density
perturbations is presently one of the most fascinating and active areas
of research, we devote to it a separate section.

\section{Gravitational waves and cosmology}

\subsection{Generation of relic gravitational waves and primordial density
perturbations}

In many situations one can neglect the non-linearity of the
gravitational field, that is, the interaction of gravitational waves 
with other gravitational fields and with themselves. However,
this is not always the case. The most dramatic example is the 
interaction of gravitational waves with the strong variable gravitational 
field of the very early Universe. A gravitational wave can be thought of as
a harmonic oscillator, while the smooth variable gravitational field of the
surrounding Universe as a gravitational pump field. The g.w. oscillator is
parametrically coupled to the gravitational pump field. This specific 
coupling is a consequence of the non-linear nature of the Einstein equations. 
The coupling provides a mechanism for the superadiabatic (parametric) 
amplification of classical waves and the quantum-mechanical generation 
of waves from their zero-point quantum oscillations \cite{gri74}. 
The word ``superadiabatic" emphasizes the fact that this effect takes place 
over and above whetever happens to the wave during very slow
(adiabatic) changes of the pump field. That is, we are interested 
in the increase of occupation numbers, rather than in the gradual 
shift of energy levels. The word ``parametric" emphasizes the mathematical
structure of the wave equation. It is a change of a parameter of the 
oscillator caused by the pump field, namely, a sufficiently rapid 
variation of its frequency, that is responsible for the considerable 
increase of energy of that oscillator. 

It is common to write the perturbed gravitational field of a 
homogeneous isotropic universe in the form:
\begin{equation}
\label{cosmetgw}
{\rm d}s^2 = a^2({\eta})[-{\rm d}\eta^2 + (\delta_{ij} + h_{ij})
{\rm d}x^i{\rm d}x^j]. 
\end{equation}
The gravitational field perturbations $h_{ij} (\eta ,{\bf x})$ can
be expanded over spatial Fourier harmonics 
$e^{\pm i{{\bf n} \cdot {\bf x}}}$, where ${\bf n}$ is a constant 
(time-independent) wave-vector, 
\begin{eqnarray}
\label{hij}
h_{ij} (\eta ,{\bf x})
= {{\cal C}\over (2\pi )^{3/2}} \int_{-\infty}^\infty d^3{\bf n}
  \sum_{s=1, 2}~{\stackrel{s}{p}}_{ij} ({\bf n})
   {1\over \sqrt{2n}}
\left[ {\stackrel{s}{h}}_n (\eta ) e^{i{\bf n}\cdot {\bf x}}~
                 {\stackrel{s}{c}}_{\bf n}
                +{\stackrel{s}{h}}_n^{\ast}(\eta) e^{-i{\bf n}\cdot {\bf x}}~
                 {\stackrel{s}{c}}_{\bf n}^{\dag}  \right].
\end{eqnarray}

The polarisation tensors ${\stackrel{s}{p}}_{ij}({\bf n}),~s=1,2$ have 
different forms, depending on whether the $h_{ij}$ represent 
gravitational waves or density perturbations. In the case of 
gravitational waves, the ${\stackrel{s}{p}}_{ij}$ 
describe the two familiar ``plus'' and ``cross'' polarisations 
introduced in Sec. 2. In the case of density perturbations, the 
polarisation tensors are:
\begin{equation} 
\label{polt}
{\stackrel{1}{p}}_{ij}({\bf n}) = \sqrt{\frac{2}{3}} \delta_{ij}, ~~~~
{\stackrel{2}{p}}_{ij}({\bf n}) = -\sqrt{3} \frac{n_i n_j}{n^2} +
\frac{1}{\sqrt{3}} \delta_{ij}.
\end{equation}
The Einstein equations for the gravitational field perturbations 
$h_{ij}$ with the polarisation tensors (\ref{polt}) can only be 
satisfied if the $h_{ij}$ are accompanied by perturbations in the 
density of matter. This is why this class of perturbations is called 
density perturbations. The difference between 
${\stackrel{s}{p}}_{ij}({\bf n})$ for, respectively, gravitational 
waves and density perturbations is responsible for the difference in 
polarisation patterns of the CMB radiation, caused by these
two classes of gravitational perturbations (see Sec. 5.2 below).

For a classical field $h_{ij}$, the quantities
${\stackrel{s}{c}}_{\bf n},~{\stackrel{s}{c}}_{\bf n}^{\dag}$ are
complex numbers. For a quantized field, they are 
annihilation and creation operators satisfying the conditions
\[
[{\stackrel{s'}{c}}_{\bf n},~{\stackrel{s}{c}}_{{\bf m}}^{\dag}]=
\delta_{s's}\delta^3({\bf n}-{\bf m})\>, \quad
{\stackrel{s}{c}}_{\bf n}|0\rangle =0 \ \ ,
\]
where $|0\rangle$ (for each mode ${\bf n}$ and $s$) is the 
initial vacuum state defined at some $\eta_0$ in the very distant past, long
before the superadiabatic regime for the given mode has started. 
The normalization constant ${\cal C}$ is determined by the requirement
that initially each mode contained only the zero-point energy 
$\frac{1}{2} \hbar \omega$. Then, ${\cal C} = \sqrt{16 \pi} l_{Pl}$ for 
gravitational waves and ${\cal C} = \sqrt{24 \pi} l_{Pl}$ for density
perturbations, where $l_{Pl} = (G \hbar/c^3)^{1/2}$ is the Planck
length. Obviously, the initial vacuum amplitude, and the entire field, should
vanish, if the Planck constant $\hbar$ is formally sent to zero.

The calculation of quantum-mechanical expectation values and correlation
functions provides the link between quantum mechanics and macroscopic
physics. Using the representation (\ref{hij}) and definitions above, one 
finds the variance of the gravitational field perturbations:
\begin{equation}
\label{hmvar}
\langle 0| h_{ij}(\eta,{\bf x}) h^{ij}(\eta,{\bf x})|0\rangle
= {{\cal C}^2\over 2\pi^2} \int_{0}^{\infty} n^2\sum_{s=1,2}
|{\stackrel{s}{h}}_n(\eta)|^2 \frac{{\rm d}n}{n}.
\end{equation}
The quantity
\begin{equation}
\label{power}
h^2(n, \eta) =
{{\cal C}^2\over 2\pi^2} n^2\sum_{s=1,2} |{\stackrel{s}{h}}_n(\eta)|^2
\end{equation}
gives the mean-square value of the metric (gravitational field) perturbations 
in a logarithmic interval of $n$ and is called the (dimensionless) power 
spectrum.  The power spectrum of metric perturbations is a
quantity of great observational importance. It defines the temporal structure
and amplitudes of the g.w. signal in the frequency bands of direct experimental
searches. It is also crucial for calculations of anisotropy and
polarisation induced in CMB by relic gravitational waves and by other
gravitational field perturations. 

To find the power spectrum at any given moment of time (for instance, today
or at the moment of decoupling of CMB from the rest of matter) we need to
know the mode functions ${\stackrel{s}{h}}_n (\eta )$ at those moments of 
time. In the case of gravitational waves, the mode functions  
${\stackrel{s}\mu}_n(\eta)$ (where
${\stackrel{s}{\mu}}_n (\eta) \equiv  a(\eta) {\stackrel{s}{h}}_n (\eta )$) 
are governed by the equation for the parametrically disturbed
oscillator \cite{gri74}:  
\begin{equation}\label{fieldeq}
{\stackrel{s}\mu}_{n}^{\prime\prime} + {\stackrel{s}\mu}_{n} \left[n^2 -
\frac{a^{\prime\prime}}{a}\right] = 0 ~. 
\end{equation}
The equation describing density perturbations in the very early Universe 
can also be reduced to the equation very similar to Eq. (\ref{fieldeq}), and 
the generating mechanism will work without change. As soon as the pump 
field (represented by the cosmological scale factor $a(\eta)$) is known, and
since the initial conditions are fully determined, the mode functions
${\stackrel{s}{h}}_n (\eta )$, as well as other properties of the 
generated fields, are unambiguously calculable. 
The important (and, strictly speaking, unknown) era of cosmological 
evolution is the stage preceeding the radiation-dominated era. We call
it an initial ($i$) era and characterize, for simplicity of calculations,
by a set of power-law scale factors: $a(\eta) =l_o |\eta|^{1+\beta}$,
where $l_o$ and $\beta$ are constants. 

The main properties of cosmological perturbations 
generated as a result of superadiabatic (parametric) amplification of 
their zero-point quantum oscillations are as follows (for more details 
see \cite{glpps}, \cite{gdb} and references there): 

1. The initial vacuum state is described by a Gaussian wavefunction. As a 
result of quantum-mechanical Schrodinger evolution, the vacuum state 
transforms into a multi-particle state known as a squeezed vacuum state. 
The distributions of amplitudes and phases acquire strongly unequal 
variances. In the general expression for the gravitational field mode, 
\[
h_{\bf n} = A_1 \sin(n\eta +\phi_1) \cos {\bf n} \cdot {\bf x} + 
A_2 \sin(n\eta +\phi_2) \sin {\bf n} \cdot {\bf x}, 
\]
the amplitudes $A_1$ and $A_2$ are drawn from a broad Gaussian distribution,
whereas the phases $\phi_1$ and $\phi_2$ are practically fixed and equal
up to $\pm \pi$. The field is a stochastic collection of standing waves and 
is characterised by a strongly modulated power spectrum. This complicated
statistical picture of generated cosmological perturbations is often 
replaced in the literature by a single word: ``Gaussian".

2. The generated gravitational field perturbations act on all
sorts of matter together. There is no reason why the 
inhomogeneities in different sorts of matter (if more than one component 
of matter was dynamically important in the very early Universe) should be 
displaced and move with respect to each other, which would constitute 
the so-called isocurvature, or entropy, perturbations. This is why the
generated density perturbations are often called ``adiabatic".

3. The primordial  spectrum (i.e. the spectrum before processing at the 
radiation-dominated and matter-dominated stages) as a function of the 
wave-number $n$ is fully determined by the variable 
pump field as a function of time $\eta$. Every interval of spectrum that
can be meaningfully approximated as a power-law function of $n$, 
was generated by an interval of a power-law evolution 
$a(\eta) \propto |\eta|^{1+\beta}$. Concretely, for the 
spectrum of Eq. (\ref{power}) one finds: $h^2(n) \propto
(l_{Pl}/l_o)^2 n^{2(\beta+2)}$. Specifically for density perturbations,
one often uses the spectral index ${\rm n}$, related to $\beta$ by
${\rm n} = 2\beta +5$. If the very early Universe was goverened by a scalar
field (the central assumption of inflationary scenaria), then, at every 
power-law interval of evolution and, hence, at every power-law interval of the
generated spectrum, there must be $\beta \le -2$ and ${\rm n} \le 1$.
The ``red" spectra $\beta < -2$ (${\rm n} < 1$) possess a serious theoretical 
difficulty: the mean-square value of the field becomes power-law divergent 
in the limit of very long waves (lower limit of integration in 
Eq.(\ref{hmvar})). The case $\beta = -2$ (${\rm n} =1$) is called the flat, 
or Harrison-Zeldovich-Peebles, or ``scale-invariant", spectrum.  

4. Gravitational waves are generated inevitably, wheras the generation of
density perturbations requires additional assumptions about the 
coupling of matter fields to gravity. The primordial (unprocessed) 
amplitudes of density perturbations can be as large as g.w. amplitudes,
but never much larger. The observed CMB anisotropy in lower multipoles 
(caused by primordial gravitational field perturbations) 
may have nothing to do with quantum mechanics, but if it does, the 
contribution of relic gravitational waves to lower multipoles must be 
substantial.    

5. The parametric mechanism is universal, and the  generation of 
primordial cosmological perturbations takes place regardless of whether 
the Universe 13 billion years later will appear to astronomers as 
spatially-flat, or not. In any case, if $\Omega_{total}$ is not 
identically 1, the present-day $\Omega_{total}$ is usually regulated ``by hand"
through the duration of the initial era. This parameter of duration does not 
affect seriously the amplitudes and spectral slopes of primordial 
perturbations.

\subsection{Detection of relic gravitational waves}

The direct detection of relic gravitational waves will be fundamentally 
important for uncovering the physics of the very early Universe. In Fig.6 we 
show the expected spectrum of today's r.m.s. amplitudes $h(\nu)$ (square
root of Eq. (\ref{power})) as a function
of frequency $\nu$ in $Hz$. The graph shows the piece-wise envelope of 
the spectrum and ignores its oscillations. Almost everything in this graph 
is the processed spectrum; the primordial part survives only at frequencies
$\nu < \nu_H$, where $\nu_H =c/ \lambda_H = H \approx 2 \times 10^{-18} Hz$ is
the Hubble frequency. This particular spectrum was derived under the 
assumption that a significant fraction of the observed large-scale CMB 
anisotropy is caused by relic gravitational waves and that the primordial 
spectral index is $\beta = -1.9$ (${\rm n} = 1.2$). This value of ${\rm n}$ 
follows from the COBE data \cite{sbg} and it has been recently reinforced, 
albeit with broad error bars, by more sophisticated analysis \cite{mbbpg}.
The evaluations of ${\rm n}$ based on larger data sets usually lead to 
smaller ${\rm n}$'s, varying in a narrow interval around ${\rm n} = 1$.
However, this precision appears to be artificial (caused by the excessively 
rigid apriori assumptions about the tested models) and the story does not 
seem to be over. In any case, we use Fig.6 for the analysis of 
detectability of relic gravitational waves. 

The r.m.s. values $h(\nu)$ are directly entering the detectability
evaluation, but they also determine the $\Omega_{gw}(\nu)$-parameter, 
which is useful for comparison of the g.w. background with other 
energy components. This parameter can be calculated according to the formula
\[
\Omega_{gw}(\nu) = \frac{\pi^2}{3} h^2(\nu) \left(\frac{\nu}{\nu_H}\right)^2.
\]
For example, one has $h(\nu) \approx 10^{-20.5},~ 
\Omega(\nu) \approx 10^{-11}$ at $\nu = 10^{-3} Hz$, and $h(\nu) 
\approx 10^{-25},~ \Omega(\nu) \approx 10^{-10}$ at $\nu = 10^{2} Hz$.
It is seen from Fig. 5 that the part of the spectrum accessible to LISA
is higher than the instrumental noise and can be measured. The ground-based
advanced interferometers are also promising. The target value of the
LIGO-II at $\nu= 10^2 Hz$ is $h_{ex}=10^{-23}$. The gap in two orders of
magnitude can be covered by cross-correlation of the outputs of two or 
more detectors. The $S/N$ will be better than 1 if the common integration
time exceeds $10^6 sec$. This does not seem to be a hopeless task.

It is possible that relic gravitational waves will be first observed
indirectly, with the help of polarisation measurements of CMB. The problem
is to distinguish the polarisation pattern caused by gravitational waves 
from that caused by density perturbations. The polarisation
arises as a result of the Thompson scattering of CMB photons on free
electrons \cite{rpzk}. To produce a net polarisation, the electrons 
should be illuminated by CMBR having a non-zero quadrupole anisotropy.
The two polarisation patterns are distinguishable if the quadrupole 
anisotropies are distinguishable. For our purposes of illustration, 
it is sufficient to consider the emission (rather than scattering) of 
electromagnetic waves by free electrons. The electrons are set in motion by, 
respectively, gravitational waves and density perturbations. 

\begin{figure}[tbh]
\centerline{\epsfxsize=7cm \epsfbox{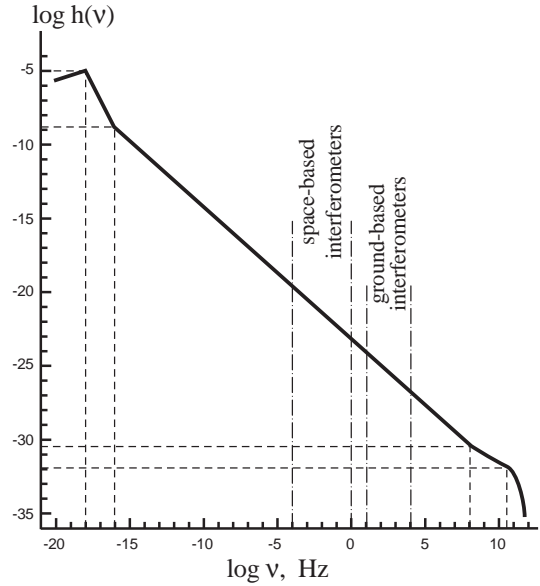}}
\caption{Expected envelope of the spectrum $h(\nu)$ for the case
$\beta = -1.9$ (${\rm n} = 1.2$)} 
\label{Fig.6}
\end{figure}

The motion of free particles in the field of a linearly-polarised 
gravitational wave is shown in Fig.1. Imagine that the moving particles 
are free electrons in the early Universe, rather than free mirrors of an 
interferometer in laboratory, that we were discussing 
in Sec.3. Then, the directions of the induced osillations of the electrons,
indicated by arrows in Fig.1, are, at the same time, the directions 
of the electric fields of the electromagnetic waves emitted by these
oscillating electrons. The pattern of arrows seen on this figure is 
the pattern of polarisation components that will be seen on the sky.

\begin{figure}[tbh]
\centerline{\epsfxsize=7cm \epsfbox{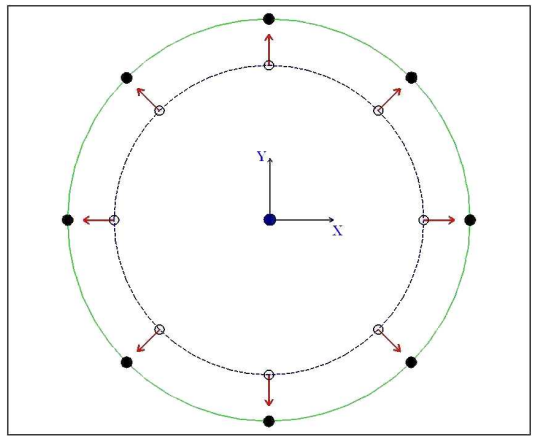}}
\caption{Motion of free particles in the gravitational field of a density
perturbation}
\label{Fig.7}
\end{figure}

The motion of free electrons induced by the gravitational field of 
a density perturbation is different. Two polarisation tensors of a density
perturbation are given by Eq. (\ref{polt}). The first one describes a
trivial purely spherically-symmetric deformation of a sphere of free 
particles. The second one, subject of our interest, describes 
the deformation which has the angular dependence of the spherical harmonic 
$Y_{lm}(\theta, \phi)$ with $l=2$ and $m= 0$, in contrast to the
combination of $Y_{lm}(\theta, \phi)$ with $l=2$ and $m= \pm 2$,  
attributable to a gravitational wave. In both cases, the polar axis $z$ is 
taken along the wave-vector ${\bf n}$ of the perturbation. 
Fig.7 shows a ring of electrons set in motion by the gravitational
field of a density mode. The pattern of these arrows is, at the same time, 
the pattern of polarisation components generated by a density perturbation. 
Comparing Fig.1 and Fig.7 one concludes that the polarisation patterns are
distinguishable, if one can study the distrubution of polarisation
Stokes parameters over a sufficiently large portion of the sky. In real 
conditions, the free electrons will be influenced by the superposition 
of many perturbation modes with arbitrary wave-vectors, but the net 
difference between these two sorts of perturbations should survive.

The gap between the expected g.w. signal and the polarisation detection 
capabilities is relatively small. Taking into account the current 
impressive activity in this area, one can hope that some decisive 
observational information about relic gravitational waves can be obtained 
pretty soon.

\subsection{Gravitational waves and inflation}

The theoretical and experimental studies of relic gravitational waves 
are endangered by absurd claims that are prevalent in 
inflationary literature. Inflationists claim that the amount of 
relic gravitational waves should be zero if
the primordial spectrum is flat, that is, if ${\rm n} =1$. There seems to
be no need to bother about relic gravitational waves, as the observations
indicate that ${\rm n}$ may indeed be close to 1. Since the ``pillars of
inflation" are popular in a part of astrophysics community, and sometimes are
said to be ``confirmed", it is important to put matters straight.  

Inflationary scenario operates with a scalar field ${\varphi}$ and
the scalar field potential $V(\varphi)$. 
Having accepted the general concept of parametrically 
amplified quantum fluctuations, inflationary theorists are performing 
their own calculations. The quantum-mechanical content of these 
calculations is usually limited to vague words, such as that 
``inflation amplified quantum fluctuations onto macroscopic scales".
Being unsure why and where the Planck constant 
$\hbar$ should enter the calculations, inflationists never write it 
explicitely; and when it is written implicitely, in the form of the 
Planck mass, $M_{Pl} = (\hbar c/G)^{1/2}$, it always stands in the wrong 
place, in the denominator of the final expression
instead of the nominator. With this sort of ``quantisation", inflationary
theorists derived their contribution to the subject of cosmological
perturbations -- the ``standard inflationary result". The ``standard
inflationary result" predicts the infinitely large amplitudes of 
today's density perturbations in the limit of the flat spectrum 
${\rm n}=1$. Indeed, one will always be able to recognise in 
inflationary papers the evaluations
relating the final ($f$) amplitudes of the perturbations to the 
initial $(i)$ values of $\varphi$ and other quantities: 
$(\delta \rho / \rho)_f \sim (h_S)_f \sim (\zeta)_f \approx (\zeta)_i \sim
(H^2/\dot \varphi)_i \sim (V^{3/2}(\varphi)/ V'(\varphi))_i 
\sim H_i/\sqrt{1 -{\rm n}}$. The denominator of the last expression is
zero for ${\rm n} =1$. The nominator is the Hubble paparameter $H_i$ at
the initial stage. $H_i$ is much larger than the Hubble parameter 
at the subsequent radiation-dominated stage, and, in any case, $H_i$ is
not zero. Therefore, the predicted final amplitudes go to infinity,
if the spectral index ${\rm n} = 1$. Inflationists are hiding this absurd 
prediction of infinitely large density perturbations by composing the 
ratio of the gravitational wave amplitude $h_T$ to the predicted 
divergent amplitude of the scalar metric perturbations $h_S$ 
(the so called ``tensor-to-scalar ratio" or ``consistency relation'': 
$h_T/h_S \approx \sqrt{1- {\rm n}}$) and declaring that it is the amount of 
gravitational waves that should be zero, or almost zero, at cosmological 
scales and, hence, down to laboratory scales. 

Certainly, the ``standard inflationary result'' is in full disagreement 
not only with the theoretical quantum mechanics, but with available 
observations too, as long as the error-boxes of the observationally 
derived spectral index ${\rm n}$ are centered at ${\rm n} \approx 1$ 
and include ${\rm n}=1$. To be consistent with inflationary predictions,
the data should not allow ``blue" spectra ${\rm n}> 1$, as the scalar field
cannot produce them, and the density amplitudes should go to infinity, 
when one processes the data assuming that ${\rm n} =1$. This spectacular 
failure of inflationary calculations is systematically painted by 
inflationists and their followers as a great success. The 
most recent example is the analysis
of WMAP data \cite{peiris}. The authors praise and follow inflationary   
derivations, and conclude that the ``tensor/scalar ratio r" is ``consistent
with zero". The ``standard inflationary result" is written in that paper 
(their formula (17)) as: 
\[
\Delta_{\cal R}^2 = \frac{V/M_{Pl}^4}{24 \pi^2 \epsilon_V}, 
\]
and the  ``tensor/scalar ratio r" (their formula (18)) as: 
\[
r= 16 \epsilon_V.
\]
Combining the second formula with the first one, one can easily see 
that if the WMAP data demonstrate that ``r is consistent with zero", 
then the WMAP data should also be consistent with an infinite numerical
value of the density 
amplitudes $\Delta_{\cal R}^2$ and, hence, with an infinite numerical
value of the induced CMB anisotropies. If the WMAP data are not 
consistent with such an infinite numerical value of  
density perturbations, then the only ``implication
for inflation" that follows from the WMAP observations is that 
the single concrete formula derived by inflationists -- their 
``standard inflationary result" -- is shown to be wrong.

It seems to the author that overenthusiasm for inflation has reached 
unscientific, even ecclesiastical proportions. 
For example, this is how inflation is characterized in
the educational programm ``Astronomy" of the Smithsonian Institutions
\cite{smith}: ``...the inflationary scenario is the best current theory of the
Universe...It has met four critical observational tests...". It is unclear
which 4 tests the inflationary school credits to itself, rather than to 
direct 
consequences of quantum mechanics and general relativity, but one can recall 
that even general relativity was characterized until quite recently as a
theory that has met only 3 critical observational tests (1 of which, 
gravitational red-shift, is not a test of specifically general relativity). 
As for professional papers, one reads almost every day claims
that the CMB and galaxy surveys are ``in spectacular agreement with an
inflationary $\Lambda$-dominated cold dark matter cosmology" (compare,
for example, with \cite{efs} and \cite{blos}). Somewhere in the text,
authors usually admit that, say, the observed quadrupole anisotropy 
is way out of the predicted value, and that the probability
of finding such a result within the ``standard" model is 
$1.5 \times 10^{-3}$ \cite{spergel}, \cite{efs}.
Surely a theory which is admitted to have only a one on a thousand chance
of being consistent with one of its crucial observational tests is not
in ``spectacular agreement" with the cosmos we are trying to understand
and should not be a subject given to self-congratulation.

\subsection{Gravitational waves and quadrupole anisotropy}

The accurate measurements of CMB by WMAP reiterate the issue of the 
gravitational wave contribution to the lower order multipoles. 
The best strategy is to rely on conclusions of general physics 
and to use the minimum number of extra hypotheses. If the general 
considerations suggest (see Sec.5.1) that the contributions of gravitational
waves and density perturbations should be of the same order of magnitude,
it is this conclusion that should be tested most thoroughly.  

\begin{figure}[tbh]
\centerline{\epsfxsize=10cm \epsfbox{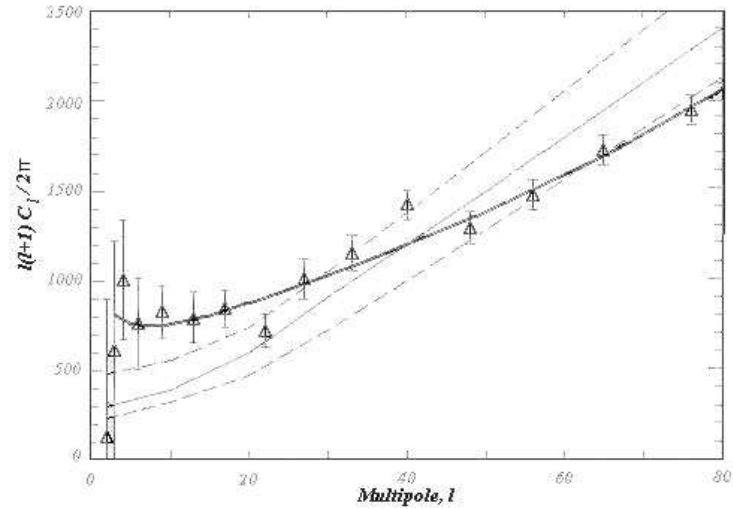}}
\caption{The WMAP data and some theoretical models}
\label{Fig.8}
\end{figure}

Fig.8 shows \cite{maus} the WMAP observational points (triangles) and the
best fit curve (solid line) of the $\Lambda$-dominated cosmology
without gravitational waves. The predicted quadrupole ($l=2$)
anisotropy is a factor of 8 higher than the actually observed value. 
The thin solid line (reproduced from \cite{gdb}, fig.12) shows the contribution 
of density perturbations alone in a model with $z_{dec} =1000,~ z_{eq} = 5000,~ 
{\rm n}=1$ that fits the position and value of the peak at $l=220$. 
This line is surrounded by the 1$\sigma$ uncertainty belt 
(shown by two dashed lines) arising due 
to the lack of ergodicity on a 2-sphere. At sufficiently large $l$'s 
the belt is approximately symmetric and its size is $\Delta C_l \approx
\sqrt{2/(2l+1)}C_l$. But the asymmetry grows towards small $l$'s, and 
specifically at $l=2$ the size of the deviation down is only $0.4$ part 
of the deviation up \cite{gm}. [The accurate displaying of this fundamental
uncertainty makes the actual quadrupole even further out of the uncertainty
belt of the $\Lambda$-dominated cosmology, than what is implied by the 
usually plotted symmetric ``cosmic variance".] Comparing the thin line
with the data points, it is difficult to avoid the conclusion (advocated
in \cite{gdb}) that in fact there exists an excess, rather than a deficit, 
of power at small multipoles, and the most natural explanation of this
excess is the anticipated contribution of gravitational waves. The increase
of the spectral index up to ${\rm n} = 1.2$ makes the agreement with
the observed quadrupole even better and implies  a somewhat larger 
amount of gravitational waves \cite{gdb}. One should remember, however, 
that all the experimental $C_l$ data, together with all the cosmological 
parameters, is a finite set of numbers. At the same time, in our hands is 
the (strictly speaking, unknown) shape of the primordial spectrum, i.e. a 
continuous function and an infinite set of numbers. A perfect 
agreement with any observed $C_l$'s and practically any set of 
cosmological parameters, not only $\Lambda =0$, can be
achieved at the expense of the properly chosen primordial spectrum.
Unfortunately, the era of ``precision cosmology" is still at some 
distance from us.  

The nature of the observed quadrupole anisotropy deserves special attention.
Most likely, it is caused by superposition of very long gravitational and
density waves. It is known \cite{gz} that a gravitational
wave produces the $Y_{2,2}$ and $Y_{2,-2}$ CMB anisotropy, whereas a density
perturbation produces the $Y_{2,0}$ CMB anisotropy. We have discussed this
distinction in Sec.5.2 in connection with the CMB polarisation. The actual
quadrupole distribution over the sky was measured by COBE \cite{kogut}.    
In Galactic coordinates,
\[
Q(\theta, \phi)= Q_1(3 \cos^2 \theta -1)/2 + Q_2 \sin 2\theta \cos \phi +
Q_3 \sin 2\theta \sin \phi + Q_4 \sin^2 \theta \cos 2\phi +  
Q_5 \sin^2 \theta \sin 2\phi,   
\]
where the measured (least noisy) components are \cite{smoot}:
\begin{equation}
\label{Qdata}
Q_1 =19.0 \pm 7.4,~~ Q_2 =2.1 \pm 2.5,~~ Q_3 = 8.9 \pm 2.0,~~ 
Q_4 = -10.4 \pm 8.0,~~ Q_5 = 11.7 \pm 7.3.
\end{equation}
Even if this $Q(\theta, \phi)$ is produced by a single gravitational wave 
or a single density perturbation, it is only in a special coordinate 
system that it can be reduced to the combination of  
$Y_{2,2}$ and $Y_{2,-2}$ or to $Y_{2,0}$, respectively.
To find out what we are dealing with, we have to build invariants, that is, 
quantities independent of the rotation of the coordinate system. One of 
invariants is  
\[
Q^2_{rms} = (4/15)[(3/4) Q_1^2 + Q_2^2 + Q_3^2 + Q_4^2 + Q_5^2],
\]
another one (see, for example, \cite{mag}) is 
\[
D = (4/5^{3/2})[(1/4)Q_1(Q_1^2 +Q_2^2/2 +Q_3^2/2 - Q_4^2 -Q_5^2) +
2Q_2 Q_3 Q_5 +Q_4(Q_2^2 -Q_3^2)].
\]
$Q_{rms}$ is always positive, wheras $D$ can be negative, but it always 
satisfies the condition $|D| \le Q_{rms}^3$. For a pure density perturbation,
$|D| = Q_{rms}^3$; and for a pure gravitational wave, $D=0$.  

Calculating the invariants and the formal errors from the data set
(\ref{Qdata}) we find
\[
Q_{rms} = (12.6 \pm 3.4) \mu K, ~~~-D^{1/3} = (6.9 \pm 12.9) \mu K
\]
As expected, the available noisy data do not allow one to prefer 
one of hypotheses over another. But, for sure, there is no indications 
whatsoever that the quadrupole anisotropy is produced by a density 
perturbation alone. If anything, the central values of $Q_{rms}$ and
$D$ indicate that the contribution of gravitational waves should be
substantial. Hopefully, the WMAP quadrupole data will be more accurate, and 
then this analysis should be repeated.

\section{Summary}

Gravitational-wave physics is a mature and at the same time a very
young science. In a sense, the relativistic gravity (general relativity)
itself is still a young science. The enormous progress in technical 
developments and  observational verifications of the theory is 
accompanied by difficult issues of its adequate description and 
interpretation, the necessity of bringing it to a closer contact with 
other branches of physics. 

It is rumored that the Nobel recognition eluded the great 
astronomer E. Hubble because of his reluctance to say about his discovery 
what the establishment wanted him to say. Apparently, the scientific 
integrity of E. Hubble allowed him to say what he believed he 
discovered -- the nonstationarity of the system of nearby 
galaxies -- whereas he was required to admit that he 
discovered the ``expansion of space". But the ``expansion of space" is
still alive and well. [For example, ``As bizarre as it may seem, space 
itself is expanding -- specifically, the vast regions of space between 
galaxies" \cite{smith}.] It is regularly proposed to be measured. The logic 
seems to be impecable. If ``space" expands by a factor of 2 in 10 billion 
years, why would not the Earth or an atom expand by 10$\%$ in 
1 billion years ? The gravitational-wave research is plagued in a similar
fashion. It is often stated that ``gravitational waves are oscillations of 
space-time itself". The next phrase seems to be logically unavoidable:
``gravitational waves act tidally, stretching and squeezing any object
that they pass through". If this phrase were correct, we would never
be able to notice gravitational waves. The device measuring, say, the 
displacements 
of free mirrors in an interferometer would be ``stretched and squeezed" 
as well. In this situation, we can probably find comfort in the wise
observation \cite{Wein}: ``I agree that much of what one reads in the
literature is absurd. Often it is a result of bad writing, rather than
bad physics. I often find that people who say silly things actually 
do correct calculations, but are careless in what they say about them." 

It seems to the author that, in the long perspective, the value of the
gravitational wave research will be in its influence on the fundamental
physics. Meanwhile, let us hope that the next gravitational-wave update
will be devoted to the fascinating nature of concrete astrophysical 
sources of gravitational waves detected by the existing instruments.

\section*{Acknowledgments}

I am grateful to S.Babak, D.Baskaran, and B.Sathyaprakash for help and
useful discussions.

\end{document}